\documentclass[manuscript]{acmart}
\usepackage[font=small]{caption}

\DeclareUnicodeCharacter{0301}{\'{e}}

\usepackage{comment}

\usepackage{etoolbox} 

\usepackage{hyperref}

\AtBeginDocument{%
  }

\setcopyright{none}
\copyrightyear{2024}
\acmYear{2024}
\acmDOI{XXXXXXX.XXXXXXX}

\acmConference[]{XXIII Brazilian Symposium on Human Factors in Computing Systems}{}{}
\acmPrice{15.00}
\acmISBN{}

\begin{document}

\title{A Comparative Study on Accessibility for Autistic Individuals with Urban Mobility Apps}


\author{Danilo Monteiro Ribeiro}
\affiliation{%
  \institution{Universidade Federal de Pernambuco}
  \city{Recife}
  \country{Brazil}
}
\email{dmr@cin.ufpe.br}

\author{Felipe Vasconcelos Melo}
\affiliation{%
    \institution{Universidade Federal de Pernambuco}
    \city{Recife}
    \country{Brazil}
}
\email{fvm3@cin.ufpe.br}

\author{Vitor Negromonte}
\affiliation{%
  \institution{Universidade Federal de Pernambuco}
  \city{Recife}
  \country{Brazil}
}
\email{vnco@cin.ufpe.br}

\author{Gabriel Walisson Matias}
\affiliation{%
    \institution{Universidade Federal de Pernambuco}
    \city{Recife}
    \country{Brazil}
}
\email{gwam@cin.ufpe.br}

\author{Adna Farias}
\affiliation{%
    \institution{Universidade Federal de Pernambuco}
    \city{Recife}
    \country{Brazil}
}

\email{alfs2@cin.ufpe.br}

\author{Celeste Azul}
\affiliation{%
  \institution{Universidade Federal de Pernambuco}
  \city{Recife}
  \country{Brazil}
}
\email{caggp@cin.ufpe.br}

\author{Ana Paula Chaves}
\affiliation{%
  \institution{Northern Arizona University}
  \city{Flagstaff}
  \country{United States of America}
}
\email{ana.chaves@nau.edu}

\author{Kiev Gama}
\affiliation{%
  \institution{Universidade Federal de Pernambuco}
  \city{Recife}
  \country{Brazil}
}
\email{kiev@cin.ufpe.br}


\renewcommand{\shorttitle}{A Comparative Study on Accessibility for Autistic Individuals with Urban Mobility Apps}
\renewcommand{\shortauthors}{D. Ribeiro, F. Melo, V. Negromonte, G. Matias, A. Farias, C. Azul, A. Chaves, K. Gama}
\renewcommand{\abstractname}{Abstract}

\begin{abstract}
Autism Spectrum Disorder (ASD) is a neurodivergent condition with a wide range of characteristics and support levels. Individuals with ASD can exhibit various combinations of traits such as difficulties in social interaction, communication, and language, alongside restricted interests and repetitive activities. Many adults with ASD live independently due to increased awareness and late diagnoses, which help them manage long-standing challenges.
Predictability, clarity, and minimized sensory stimuli are crucial for the daily comfort of autistic individuals. In mobile applications, autistic users face significant cognitive overload compared to neurotypicals, resulting in higher effort and time to complete tasks. Urban mobility apps, essential for daily routines, often overlook the needs of autistic users, leading to cognitive overload issues. This study investigates the accessibility of urban mobility apps for autistic individuals using the Interfaces Accessibility Guide for Autism (GAIA). By evaluating various apps, we have identified a common gap regarding accessibility for people with Autism Spectrum Disorder (ASD). This limitation relates to the absence of a functionality that allows users on the autism spectrum to customize the characteristics of the textual and visual elements of the software, such as changing the text font, altering the font type, and adjusting text colors, as well as native audio guidance within the applications themselves. Currently, the only function in this context is for visually impaired people, which completely changes the user experience in terms of navigation.
\end{abstract}



\keywords{Acessibility, Autism, Urban Mobility Apps, Smart Cities}


\maketitle

\section{Introduction}

Autism Spectrum Disorder (ASD) is a neurodivergent condition that encompasses a wide range of characteristics and levels of support. As a spectrum, individuals with ASD can exhibit various traits, with combinations that may vary from one person to another. These traits include difficulties in social interaction, communication, and language, along with restricted interests, repetitive behavior patterns, and challenges in understanding concepts with ambiguous interpretations~\cite{american2013diagnostic,moresi2018tecnologia,opastea}.
Although ASD involves a wide range of symptoms and levels of severity, many adults lead independent and functional lives~\cite{howlin2017autism}. With advancements in understanding and knowledge of ASD, a recent trend is the increasing number of adults who, despite never considering themselves autistic, are now being diagnosed with ASD~\cite{happe2016demographic}. This has led to a better understanding of the difficulties they have always lived with~\cite{lupindo2023late}.

In general, for many autistic individuals, predictability, clarity, and minimizing sensory stimuli are essential elements for a more comfortable and secure daily life \cite{green2018sensory,sinha2014autism}. 
When applied to the context of mobile applications, autistic people face greater challenges of cognitive overload and have less cognitive control compared to neurotypical individuals. This results in greater effort and time when using apps to complete the same tasks~\cite{mackie2016reduced,yaneva2019adults}. 

The daily use of urban mobility apps is part of our routines, with a wide range of apps varying according to the mode of transportation~\cite{silva2023smart}. However, autistic individuals are often a neglected user group in this context, as they may face unique challenges related to mobility and interaction with the urban environment~\cite{araujo2011transporte, cerqueira2021autismo}. Cognitive overload is a frequent issue with these types of apps, which can cause difficulties for users with ASD~\cite{mauro2022role}.
Some research~
\cite{TalitaEdnaldoGAIAteoria,aguiar2022autismguide} has developed heuristics to investigate software accessibility in the context of autistic individuals. Among them, the Accessibility Guide for Autism Interfaces (GAIA) aims to establish design guidelines and recommendations that promote accessibility~
\cite{gaiaexplain,TalitaEdnaldoGAIAteoria}, serving as an essential resource for software developers and digital educators in creating websites adapted to the needs of individuals with ASD. However, the use of these guides is generally targeted at apps specifically designed for autistic children~\cite{santiago2022user}. The current literature does not present, as far as could be found, studies that assess the accessibility of autistic users in the context of urban mobility apps. In a recent study on mobility apps~\cite{Callil2021}, funded by an urban mobility app, accessibility aspects—especially concerning autistic individuals—are not considered.

Given the importance of GAIA and the context of urban mobility, this article aims to investigate the accessibility of mobile apps geared toward urban mobility for autistic individuals. To do so, in the context of an urban mobility project focused on accessibility for neurodivergent individuals, we used GAIA's heuristics to evaluate popular urban mobility apps. The study evaluated the interfaces of popular urban mbility apps used in Brazil (Google Maps, Waze, Moovit, Cittamobi, Uber, 99, Bike Itaú, and Tembici) 
and their adequacy to the needs of autistic individuals, with the goal of identifying gaps and opportunities for improvement in the applications. Through a critical analysis supported by GAIA, we aim to provide valuable insights for developers and designers to guide them in creating more inclusive and accessible solutions for users with ASD.

The theme presented here intersects at least two major research challenges in HCI in Brazil~\cite{baranauskas2012grandihc}: (1) The Future, Smart Cities, and Sustainability; and (2) Accessibility and Digital Inclusion. The first is related to the theme of urban mobility, and the second is because we are addressing a group that, despite not being explicitly included in the original text of the challenge, is considered to have a disability in Brazil, according to the law that establishes the National Policy for the Protection of the Rights of Persons with Autism Spectrum Disorder~\cite{Lei2012}.

\section{Background}
\subsection{Teoria da Carga Cognitiva}

John Sweller (2001) developed a theory that seeks to elucidate the influence of memory in the learning process \cite{sweller2011cognitive}. To this end, he highlights that working memory, which is responsible for processing information in real time, has a limited capacity to handle an extensive amount of information \cite{alves2017dimensoes}. In this context, when a task demands the use of a large number of cognitive resources, memory can become overloaded, resulting in impairments to the learning process \cite{sweller2011cognitive}. Given this scenario, Sweller categorizes cognitive load into two distinct components \cite{sweller1994some}:
\begin{enumerate}
    \item Intrinsic Cognitive Load (ICL): Refers to the cognitive load inherent to the very nature of the task, determined by the intrinsic complexity of the task itself.
    \item Extraneous Cognitive Load (ECL): Represents the additional cognitive load imposed on the task by the way it is presented to the learner. This load is influenced by the organization of information and the presence of distractions in the learning environment.
\end{enumerate}

Sweller suggests an ideal distribution for these two cognitive loads, advocating for a higher intrinsic cognitive load in contrast to a reduced extraneous cognitive load \cite{sweller1994some}. This configuration aims to provide a favorable condition for the effective processing of received information. The reduction of extraneous cognitive load allows the intrinsic cognitive load, which demands continuous use of cognitive resources, to receive more focused attention. This approach promotes improved information processing, consequently resulting in more effective learning by the individual.

\subsection{Autism Spectrum Disorder}
Autism Spectrum Disorder (ASD) is a developmental disorder characterized by dysfunctions in communication, social interaction, and repetitive patterns of behavior, interests, or activities~\cite{american2013diagnostic}. Individuals with ASD face various challenges related to the use of Executive Functions (EFs)~\cite{hovik2017distinct,parhoon2022psychometric,tamm2014open}, which affect their daily lives, interpersonal relationships, and professional and academic experiences. EFs, also known as executive control or cognitive control, are a set of functions essential for directing and regulating various intellectual, emotional, and social skills, used to perform tasks, solve problems, or face new situations~\cite{lezak2004neuropsychological,diamond2022executive,american2013diagnostic}. EFs involve planning and organization, decision-making, reasoning, and problem-solving activities. Among these functions, working memory, inhibitory control, and cognitive flexibility are considered core functions~\cite{diamond2020biological,diamond2022executive}. 

Studies indicate that individuals with ASD exhibit general deficiencies in working memory~\cite{wang2017meta}, which is the executive function directly linked to cognitive load theory~\cite{sweller2011cognitive}. ASD is classified into three levels: Level 1, where the individual requires minimal support and typically presents with social and communication difficulties; Level 2, which requires substantial support due to more pronounced challenges in social interaction and restrictive behaviors; and Level 3, where very substantial support is necessary due to severe communication difficulties and behaviors that significantly interfere with daily functioning~\cite{american2013diagnostic}. Individuals diagnosed at Levels 1 and 2, who generally require less intensive support, are the autistic individuals most present in the workforce, but their needs are often misunderstood ~\cite{ortega2009deficiencia}.

Currently, many of the tasks performed by autistic people on a daily basis involve the use of mobile applications. The main difficulties this population may encounter, for example, in navigating websites~\cite{TalitaEdnaldoGAIAteoria}, are in the areas of reading, verbal and/or linguistic comprehension, visual comprehension, focus, and attention, among others. The lack of accessibility resources that reduce cognitive load when using these applications impacts the daily lives of autistic people. The amount of material to instruct developers in the planning and implementation of solutions adapted to the needs of people with ASD is still scarce ~\cite{TalitaEdnaldoGAIAteoria}.

Although there are solutions developed to increase accessibility and enhance the abilities of people with autism, the production of these solutions is largely concentrated in the scientific field and lacks practical applicability. One of the activities supported by mobile applications that autistic people frequently engage in is urban mobility, which will be described in the next section.

\subsection{Urban Mobility Apps}

Urban mobility encompasses everything from walking (short distances) to traveling in vehicles (longer distances), which over the years has evolved to include various modes of transportation, such as bicycles, trams, trains, buses, and ridesharing,to meet the diverse transportation needs in urban areas ~\cite{alessandretti2022multimodal}. In a smart city strategy, Smart Mobility (SM) is a fundamental dimension that uses technology to facilitate people's movement through cities via different modes of transportation~\cite{silva2023smart}. 

Some studies go beyond the concept of SM and introduce the idea of **Mobility as a Service (MaaS)**~\cite{li2017mobility,ho2024mobility}. This perspective envisions the integration of various modes of transportation through a unified platform that would allow the use of a single app or integrated platform. In a recent study~\cite{ho2024mobility}, MaaS is seen as having great potential to transform transportation, especially in developing countries. However, the idea of an integrated platform has been overshadowed by the view of MaaS as a complement to individual vehicle transportation, due to obstacles such as a lack of collaboration between secretariats and government departments, leading to inefficiencies and duplicated efforts~\cite{faria2017smart}. What is currently seen in various scenarios in developing countries, particularly the reality in the Brazilian context, is a set of various apps (e.g., Uber, Google Maps, Waze) from different companies that operate in isolation and without direct integration, lacking uniformity in interfaces or user experience, which can cause cognitive overload for users.

\subsection{Cognitive Overload in Apps}
In general, an app with cognitive overload elicits negative feelings in its users, who prefer applications with lower cognitive load (e.g., less data input, a limited number of available features to choose from)\cite{szinay2021perceptions}. This cognitive overload is also a problem in urban mobility apps\cite{mauro2022role}, as it impairs decision-making skills and can even lead users to avoid using the apps altogether.

For people with autism, the issue of cognitive overload is even more concerning. Compared to neurotypical individuals, they may have less cognitive control and a reduced ability to cope with time constraints~\cite{mackie2016reduced}. In the context of app usage, individuals on the autism spectrum generally take more time and expend more cognitive effort than neurotypical individuals to complete the same tasks~\cite{yaneva2019adults}.

Therefore, greater attention must be given to user interface design, as a less intuitive interface can increase cognitive load and make it harder to understand the necessary actions within the app~\cite{tarouco2006estrategias}. For example, Apple has been implementing accessibility features to enhance inclusion in the use of its devices~\footnote{https://www.apple.com/br/newsroom/2023/05/apple-previews-live-speech-personal-voice-and-more-new-accessibility-features/}. Among the offered features, the most important for autistic individuals include Assistive Access, Live Speech, and Point and Speak. Assistive Access is aimed at individuals with cognitive disabilities, concentrating the most-used activities within apps, simplifying the experience. Live Speech allows people with speech difficulties to type their responses during phone calls. Finally, Point and Speak facilitates interaction with objects in the environment. For example, if a person points their phone's camera at a microwave and indicates a specific button, the system will speak the label of that button.

In addition to these, other initiatives aim to reduce the cognitive load of apps for autistic individuals. Among them is GAIA, a guide for building websites suitable for autistic users, which will be presented next.

\section{Recommendations for Developing More Accessible Software}

Guides such as AutismGuide, COGA, and GAIA provide recommendations for software development aimed at autistic individuals. In AutismGuide ~\cite{aguiar2022autismguide}, the authors conducted a systematic literature review to find recommendations for developing accessible applications for the autistic population. At the end of the study, the authors proposed a set of 69 recommendations in the following categories: general usability, non-functional requirements, functional requirements, adaptability, direction, visual load, compatibility, explicit control, explicit codes, error management, and consistency.

The **Cognitive and Learning Disabilities Accessibility Task Force (COGA)** is another tool designed to make the web accessible to people with cognitive and learning disabilities, including but not limited to autism. In the context of autism, COGA suggests 76 web accessibility recommendations, organized into 10 categories: (I) Personalization and User Preferences; (II) Security and Privacy on the Web; (III) Multimodal Content – Video; (IV) Multimodal Content – Text; (V) Distractions; (VI) Voice Systems; (VII) Online Payments; (VIII) Navigation, Layout and Multimedia; and (X) Online Security.

Specifically designed to evaluate interfaces from the perspective of autistic users, **GAIA (Guidelines for Accessible Interfaces for People with Autism)** is a set of recommendations that can serve as a manual for interface development~\cite{gaiaexplain}. The guide was developed with the aim of promoting accessibility for individuals with ASD, with the goal of complementing existing materials on accessibility for neurodivergent individuals, defining the parameters of what works or doesn't work for autistic individuals, building a repository of best practices and guidelines, and ultimately supporting developers in designing inclusive interfaces~\cite{gaiaexplain}. GAIA is divided into ten modules, with each module containing a set of guidelines. These modules aim to provide a wide range of heuristic tools to guide the development of websites in an inclusive, simple, and accessible manner~\cite{TalitaEdnaldoGAIAteoria}. According to the authors, the modules are distributed as follows:
\begin{itemize}
    \item VVT - Visual and Textual Vocabulary: addresses the appropriate use of texts and images, considering the particular needs of people with autism.
    \item CTM - Customization: provides guidelines related to functions that allow users to adapt interfaces to their needs.
    \item EGM - Engagement: refers to how to enable users with ASD to focus and pay more attention when using a system, allowing them to interact and engage better with the system.
    \item RR - Redundant Representations: alongside multimedia guidelines, redundant representations refer to guidelines that ensure that information is not tied exclusively to one presentation format (text, audio, or video).
    \item MT - Multimedia: refers to the appropriate use of multimedia resources in web interfaces to work on memory, attention, visual and textual comprehension, and sensory integration.
    \item VES - System Status Visibility: informs the progress of tasks performed by the user, including clearly providing information about errors, help instructions, and information related to changes in the state of elements.
    \item RP - Recognition and Predictability: addresses issues related to the design of interface elements that clearly identify how they work without deep investigation or high cognitive effort.
    \item NVG - Navigability: concerns the structure of navigation between web pages. Large amounts of information and links can make hypermedia usage an overwhelming experience for individuals with ASD.
    \item RA - Response to Actions: providing feedback on actions performed in the interface is a common usability recommendation regardless of the characteristics of the users. However, incomplete feedback or the absence of feedback is critical for autistic individuals, particularly children, due to their common difficulties with retaining attention, handling changes, and understanding verbal instructions.
    \item ITST - Touchscreen Interaction: presents recommendations for the use of touchscreen interfaces, as websites and web applications are increasingly being accessed through mobile devices.
\end{itemize}

There are already several studies using GAIA to evaluate accessibility for autistic individuals. For example, Santiago and Marques ~\cite{santiago2023exploring} investigated usability complaints related to autistic children in app reviews on the Play Store, where recurring complaints about system status visibility and customization were identified. Mendonça et al. ~\cite{de2018redesign} conducted a study aimed at redesigning apps for autistic children using GAIA recommendations. Mendonça and Marques (2019) ~\cite{mendoncca2019experiencia} also conducted an investigation of the YouTube app to verify whether it had accessibility features for autistic children. They observed that 1) there is no error prevention for button touch sensitivity, 2) the app does not allow text font adjustment, and 3) some buttons do not appear to be clickable.

Mendes ~\cite{mendes2023usabilidade} developed a study to investigate the accessibility of messaging apps and proposed a Likert-scale questionnaire for evaluating mobile app accessibility based on GAIA. The questionnaire contains over 80 items across the 10 categories proposed by Britto \cite{TalitaEdnaldoGAIAteoria}.

The works cited in this section highlight the success of using GAIA in app evaluations. However, it is important to note that, to the best of our knowledge, there are no studies applying these interface evaluation technologies in the context of urban mobility, particularly for autistic adults.

\section{Method}

This research aims to analyze the accessibility of urban mobility apps from the perspective of autistic people, conducting a comparative study among the main commercial apps through the lens of the GAIA framework.

The study is primarily exploratory, as its goal is to gain familiarity and understand the accessibility of these apps in order to generate new research hypotheses~\cite{gil2015tipo}. This research is also descriptive, as it seeks to describe the characteristics of a family of apps~\cite{gil2015tipo}.

\subsection{Comparative Method}

We utilized the comparative method to investigate the functionalities of urban mobility apps. Schneider and Schmitt~\cite{schneider1998uso} state that the comparative method aims to investigate phenomena or facts in order to highlight differences and similarities between them. There are two major stages in the comparative method~\cite{schneider1998uso}: an analogical stage, which seeks to identify similarities between the cases being studied, and a contrasting stage, where the differences between the cases are analyzed. Three steps are suggested for conducting studies using the comparative method:
\begin{itemize}
    \item The selection of two or more comparable cases/phenomena: with the intent of covering the main modes of transport and apps available on Google Play Store in Brazil, four urban mobility categories were selected, namely maps, buses, ridesharing,and bicycles; and two mobile apps were chosen in each category. The selection of the apps was based on the number of downloads in their category, resulting in the selection of the most popular ones. The selection occurred in September 2023.
    \item The definition of the elements to be compared: the set of heuristics defined by GAIA~\cite{TalitaEdnaldoGAIAteoria} was used to identify whether or not the app met the criteria.
    \item Generalization: based on the evaluation results for each app, aggregated by mode of transport, this research presents a compilation of information to generate new hypotheses and research opportunities in the context of urban mobility apps for autistic users.
\end{itemize}

\subsection{Procedures and Tools}\label{sec:Analise}

The research team was composed of three senior researchers (PhDs in Computer Science) working in the fields of Software Engineering and Human-Computer Interaction, along with four undergraduate students (Information Systems and Computer Engineering) involved in a scientific initiation project focused on developing accessibility strategies for neurodivergent users in urban mobility apps. The evaluation was conducted by the four students, referred to as evaluators throughout the text, led by one of the senior researchers, who is also active in the software development industry.

GAIA was chosen as the tool for conducting the analysis process because it was designed to assess and design technologies for autistic users~\cite{TalitaEdnaldoGAIAteoria}. Although other guides such as the AutismGuide \cite{aguiar2022autismguide} and COGA \cite{w3c_wcag_coga} exist, only GAIA included a category specifically aimed at evaluating touchscreens, which is one of the main features of urban mobility apps.

The data analysis process in this research occurred in three stages. In the first stage, the evaluators were introduced to GAIA and related works. Subsequently, the authors of the research selected a mobile app from a major Brazilian delivery company for evaluation. In this stage, a senior researcher, along with the four evaluators, conducted an analysis of the GAIA heuristics to outline the evaluation process. The primary goal of this stage was to discuss possible differences in the interpretation of each evaluation item and to establish criteria for evidence collection and the overall execution process. This step aimed to align the evaluators' understanding and ensure they had the same starting point for analyzing the apps.

Once consensus was reached in the joint analysis, the four evaluators individually proceeded to evaluate two apps in each urban mobility category (i.e., maps, ridesharing,bicycles, and buses). In other words, each evaluator was responsible for conducting two evaluations in one category. As a result, the research encompasses a total of eight evaluations (two apps in each category). In cases of uncertainty regarding an item, the evaluators discussed the specific item and made a consensus decision. If no consensus was reached, the senior researchers participated in the decision-making process. Weekly meetings were held to track progress and ensure convergence.

Using the GAIA heuristics proposed by~\cite{TalitaEdnaldoGAIAteoria}, each evaluator performed the evaluation of two apps in the same category. For instance, one of the heuristics is: "Provide clear instructions that help users gain an overview of the content and guide them, such as warning boxes, tables of contents for long texts, or instructions below interactive elements." The evaluator, therefore, checked whether the app met the heuristic, and if so, the app earned a point in the corresponding GAIA module. At the end of the process, the results from all evaluators were compiled to create an accessibility score, as described in section \ref{tab:resultados}.

The apps were installed on a smartphone running Android, as in the year the evaluation was conducted, Android had more than 78\% of the mobile operating system market share in Brazil, while iOS had a 21.5\% share, according to studies by Statista and Statcounter\footnote{https://gs.statcounter.com/os-market-share/mobile/brazil}.

\subsection{Ethical Considerations}

Due to the absence of data collection involving human subjects (e.g., interviews, questionnaires), this research did not require approval from an ethics committee. The entire research process and data collection for the app analysis were conducted by the researchers, who are the authors of this article.

\section{Results}

Table \ref{tab:resultados} presents a summary of the results from the heuristic analysis. In the table, each row represents a GAIA module. The value in parentheses indicates the maximum score an app could achieve in the respective module. The columns contain the apps grouped by category, and each cell indicates the score obtained by the app in each GAIA module. The full table of results is available for open access\footnote{\href{https://figshare.com/s/8ee278ab50df3433839d}{https://figshare.com/s/8ee278ab50df3433839d}}.

\begin{table}[]\small
\caption{Analysis of attributes found in the apps according to GAIA}\label{tab:resultados}
\begin{tabular}{lrrrrrrrr}
\textbf{Category}                    & \multicolumn{2}{c}{\textbf{Maps}}    & \multicolumn{2}{c}{\textbf{Buses}}   & \multicolumn{2}{c}{\textbf{Ridesharing}} & \multicolumn{2}{c}{\textbf{Bicycles}} \\
\textbf{App}                         & \textbf{Google Maps} & \textbf{Waze} & \textbf{Moovit} & \textbf{Cittamobi} & \textbf{Uber}        & \textbf{99}       & \textbf{Bike Itaú} & \textbf{Tembici} \\
\textbf{GAIA module (maximum score)} & \textbf{}            & \textbf{}     & \textbf{}       & \textbf{}          & \textbf{}            & \textbf{}         & \textbf{}          & \textbf{}        \\
\hline
Visual and Textual Vocabulary (13)   & 13                   & 13            & 12              & 12                 & 13                   & 10                & 12                 & 12               \\
Customization (15)                   & 6                    & 6             & 2               & 1                  & 2                    & 1                 & 1                  & 1                \\
Engagement (13)                      & 10                   & 12            & 9               & 10                 & 10                   & 5                 & 9                  & 9                \\
Redundant Representations (10)       & 8                    & 7             & 4               & 4                  & 7                    & 7                 & 2                  & 2                \\
Multimedia (6)                       & 6                    & 4             & 2               & 3                  & 4                    & 4                 & 4                  & 4                \\
System Status Visibility (7)         & 6                    & 6             & 4               & 6                  & 6                    & 6                 & 1                  & 1                \\
Recognition and Predictability (11   & 10                   & 9             & 8               & 8                  & 10                   & 8                 & 8                  & 8                \\
Navigability (7)                     & 6                    & 7             & 7               & 6                  & 7                    & 6                 & 1                  & 1                \\
Responses to Actions (4)               & 3                    & 3             & 0               & 0                  & 1                    & 1                 & 1                  & 1                \\
Touchscreen Interaction (2)          & 1                    & 1             & 1               & 1                  & 2                    & 2                 & 1                  & 1                \\
Total score (88)                     & 69                   & 68            & 49              & 51                 & 62                   & 50                & 40                 & 40              
\end{tabular}
\end{table}

\begin{figure*}
    \centering
    \includegraphics[width=0.65\linewidth]{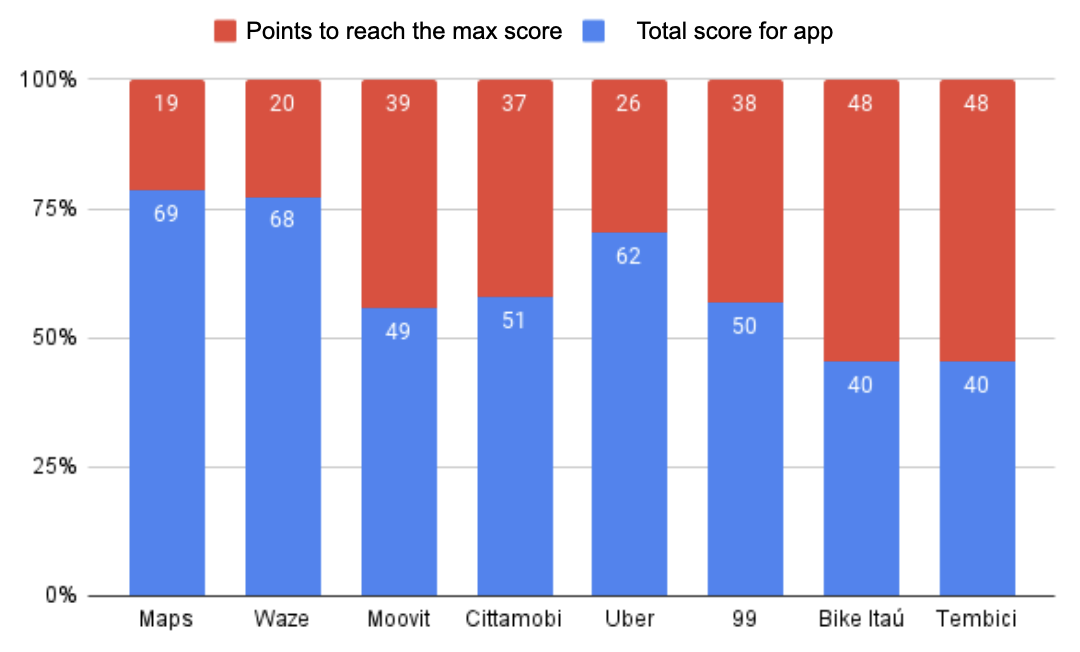}
    \caption{Illustration of the percentage of possible points versus the percentage of points found in the evaluation of the respective application.}
    \label{fig:totalxobtido}
\end{figure*}

Figure \ref{fig:totalxobtido} presents a perspective of the percentage of points evaluated in GAIA, considering the total points obtained by the apps. The score indicating compliance with GAIA heuristics is shown in blue. It is possible to observe that the map apps (i.e., Maps and Waze) have the best accessibility for autistic individuals according to GAIA. The category with the lowest compliance score is the bicycle apps, which scored only 40

The following sections present the results for each category individually

\subsection{Maps}

\begin{figure*}[h]
    \centering
    \includegraphics[width=1\linewidth]{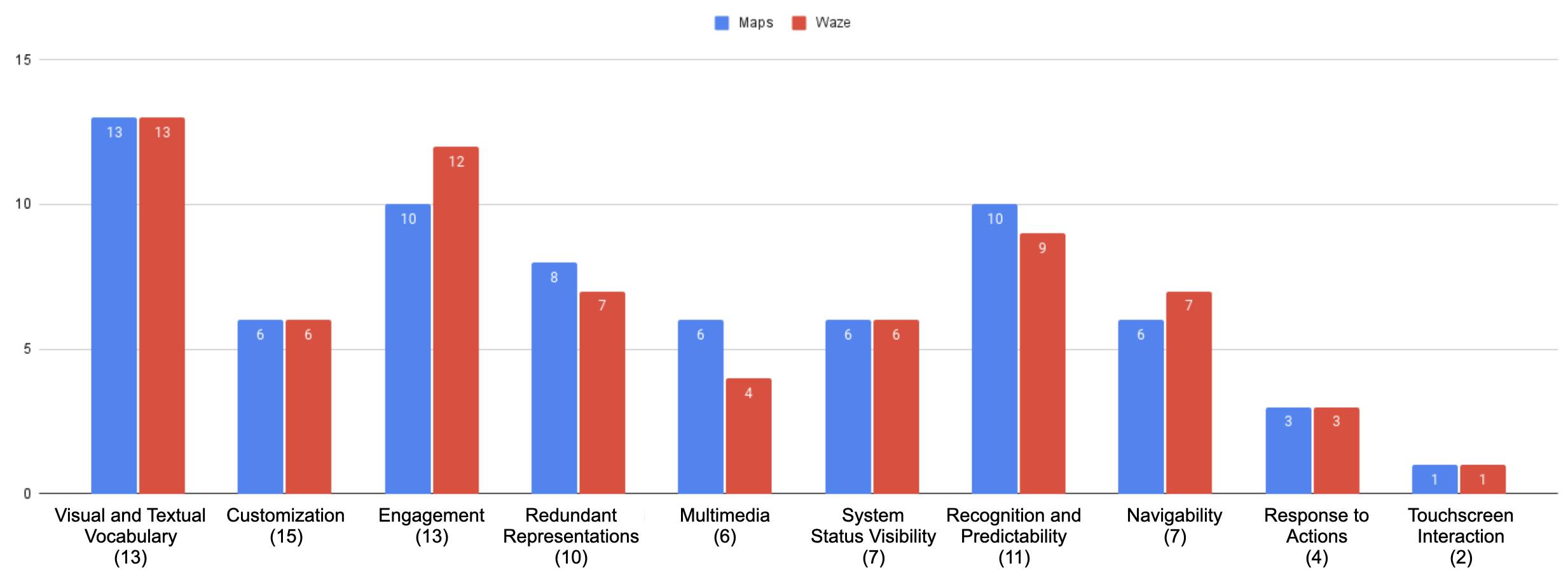}
    \caption{Analysis of the Maps category}
    \label{fig:total_mapas}
\end{figure*}

Figure \ref{fig:total_mapas} shows an overall analysis of the ``Maps'' category, comparing Google Maps (version 6.94), in blue, and Waze (version 4.99), in red, considering each GAIA module. The number in parentheses next to the module name indicates the total possible points in that respective module.

As illustrated, both apps have a significant number of features related to usability and accessibility. However, both apps scored only 6/15 points in the Customization category. The ability to customize the number of elements on the screen, reduce distractions such as animations and non-conventional fonts, and provide feedback after user interactions are critical factors for users with ASD, and the absence of these elements may compromise the experience of those using these apps.

\begin{figure}[h]  
    \centering
    \includegraphics[width=0.5\textwidth]{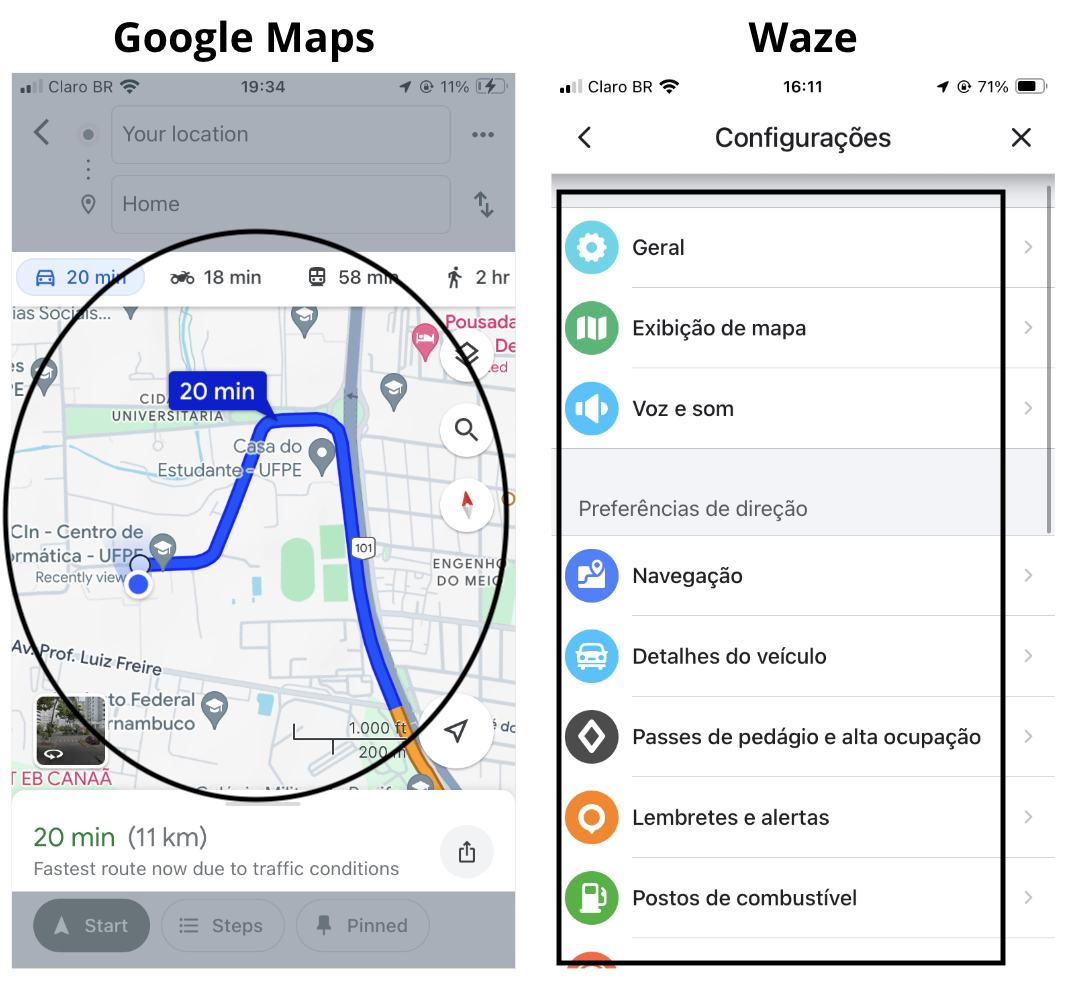}
    \caption{Issues in the submodule (Flexible Interfaces) in the Customization category and in the submodules (Eliminate Distractions) and (Visual Organization) in the Engagement category}
    \label{fig:negativo-mapas}
\end{figure}

\begin{figure}[h]  
    \centering
    \includegraphics[width=0.4\textwidth]{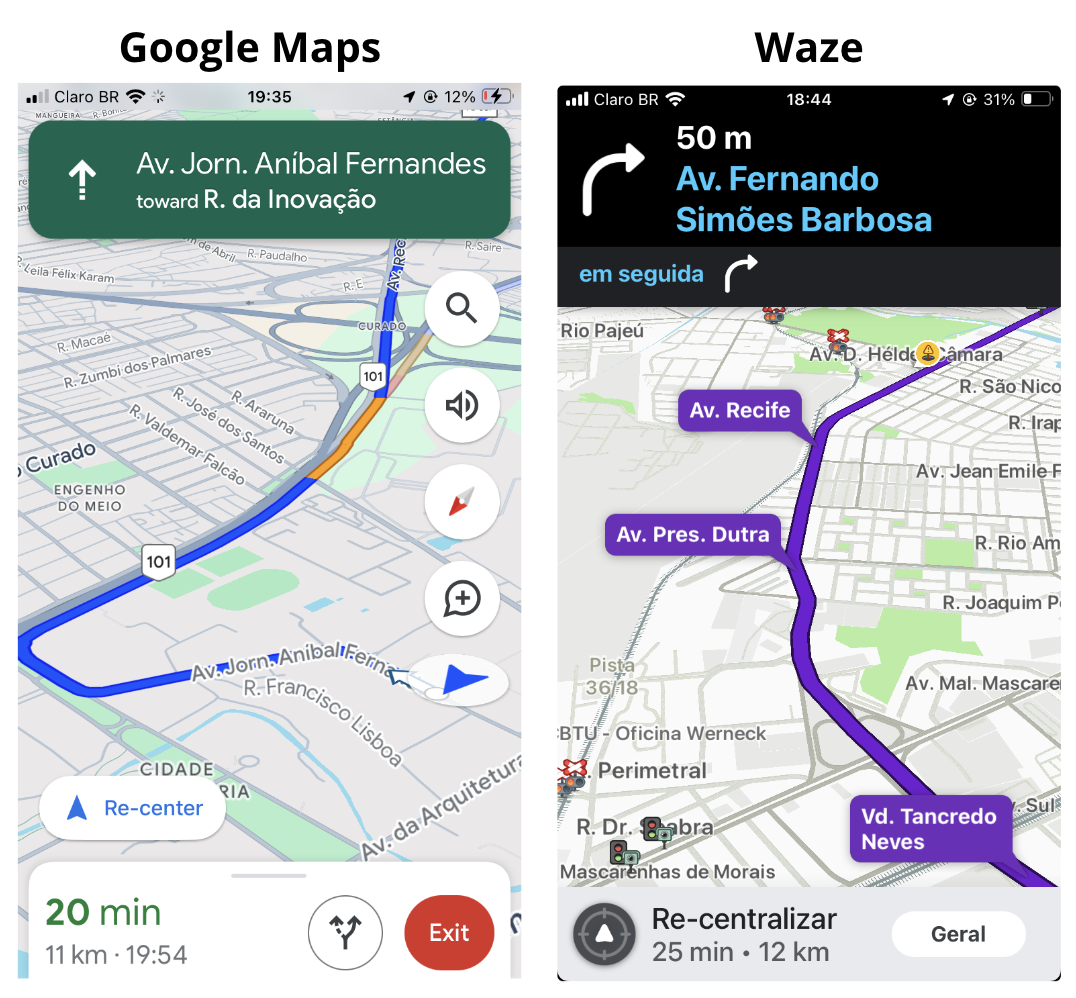}
    \caption{Positive points: Colors, Texts, Readability, Simple Navigation}
    \end{figure}

Regarding the positive points of the apps, both have a color palette that helps distinguish primary elements from secondary ones. Additionally, when the user starts the navigation process by highlighting the route to the destination, this accounts for 10/13 (Maps) and 12/13 (Waze) possible points in the Engagement category of GAIA, as shown in Figure~\ref{fig:negativo-mapas}. Furthermore, both apps present an appropriate layout of elements on the screen.

\subsection{Bicycles}

In the ``Bicycles'' category, after analyzing the apps "ItauBike" (version 9.8.5) and "TemBici" (version 9.8.5), it was observed that both present significant gaps in terms of accessibility for autistic users. Although they show some compliance with GAIA heuristics, both apps scored approximately 45\% of the possible points, highlighting the need for improvements to ensure a more inclusive and user-friendly experience for individuals with ASD.

\begin{figure*}[h]
    \centering
    \includegraphics[width=1\linewidth]{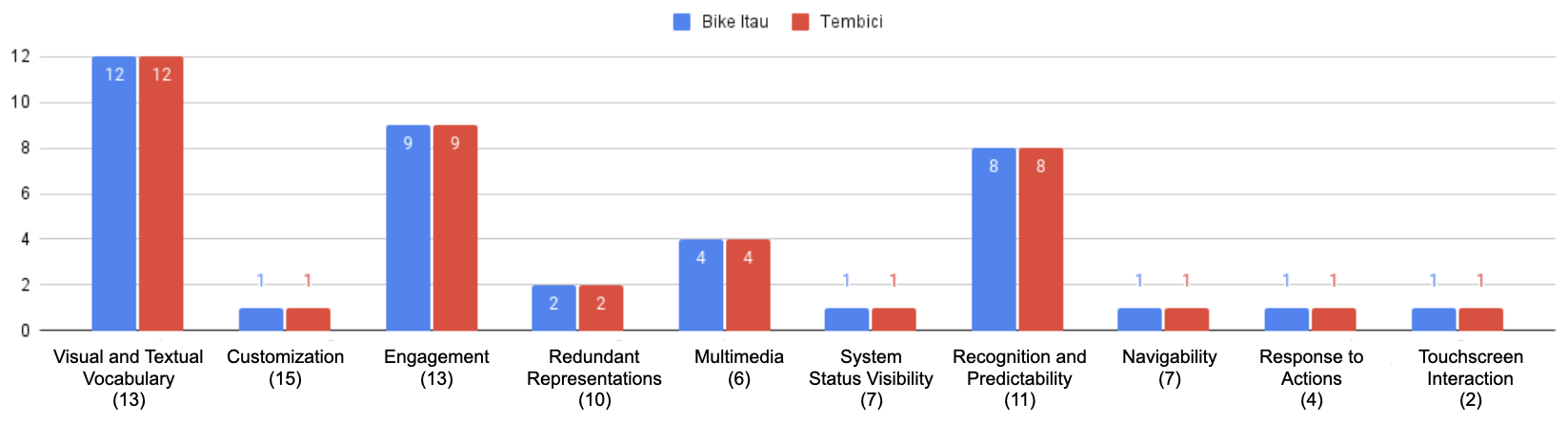}
    \caption{Analyis of the Bicycle category}
    \label{fig:total-bicleta}
\end{figure*}

Figure \ref{fig:total-bicleta} presents the overall analysis of the "Bicycles" category. Two modules stand out for their low compliance with the evaluated heuristics: Customization and Redundant Representation.

The main customization issues are related to the lack of essential accessibility features to meet the needs of autistic users, such as font and size adjustments, more varied color options, and audio guidance.

A possible explanation for the absence of these features is the reliance on the phone’s native settings for configuring these options, which are usually offered for visually impaired users. However, not only are these functions not specifically designed for the neurocognitive needs of a person with ASD, but navigating the phone’s native settings may also be challenging for these users. Additionally, native customization may not be well-suited to the app in use, increasing the user's discomfort with the interface. Therefore, it is important to offer app-specific customization features, allowing autistic users to adapt the appearance of apps according to their individual preferences.

Both apps in the "Bicycles" category failed in the areas related to the Visual (G05) and Flexible Interfaces (G07) modules. The apps need to divide the screen into sections. Moreover, it is necessary to incorporate features that facilitate contextual interpretation and interaction for users with varying degrees of autism, such as using emojis to express emotions.

The strengths of the apps include a minimalist interface with few screens and elements, as well as simplified navigation features, providing a smooth experience in terms of speed and practicality. Elements such as the use of colors to indicate movement and feedback when completing a task contribute to this experience.

\begin{figure}[h]  
    \centering
    \includegraphics[width=0.45\textwidth]{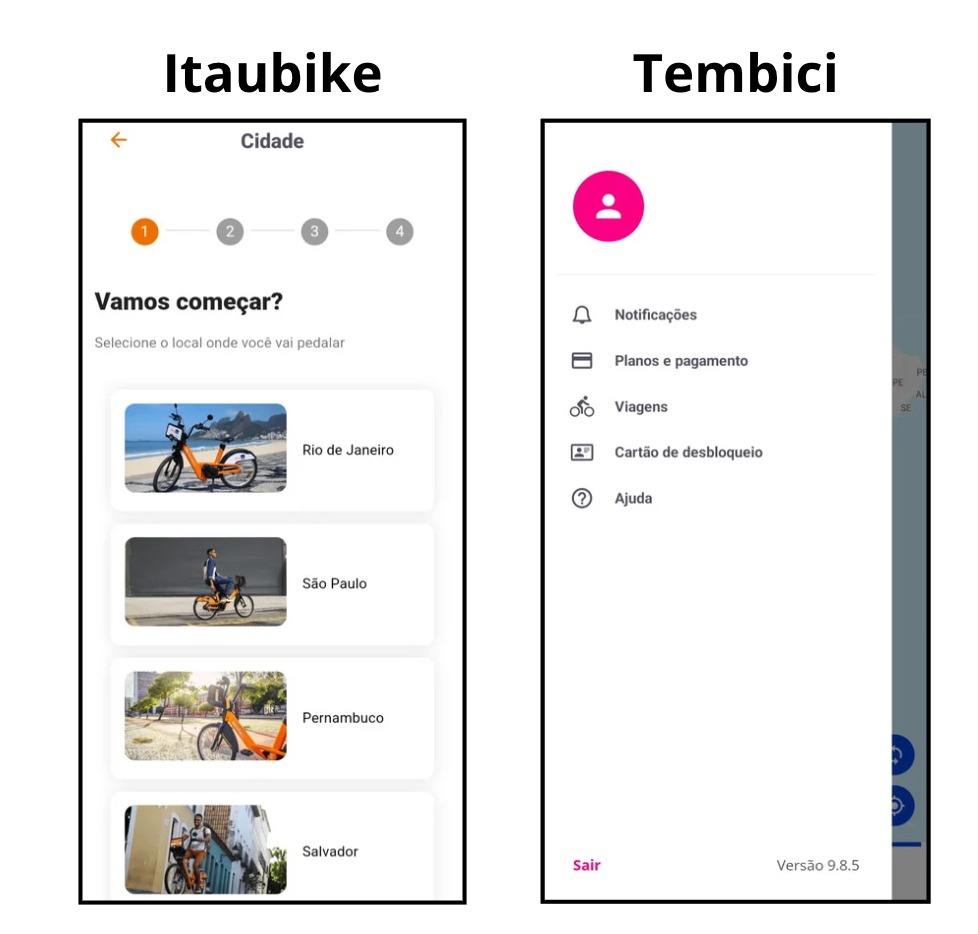}
    \caption{Positive example: Recognition and Predictability Module and the Visual and Textual Vocabulary submodule.}
    \label{front do tembici and bikeitau}
\end{figure}

\begin{figure}[h]  
    \centering
    \includegraphics[width=0.45\textwidth]{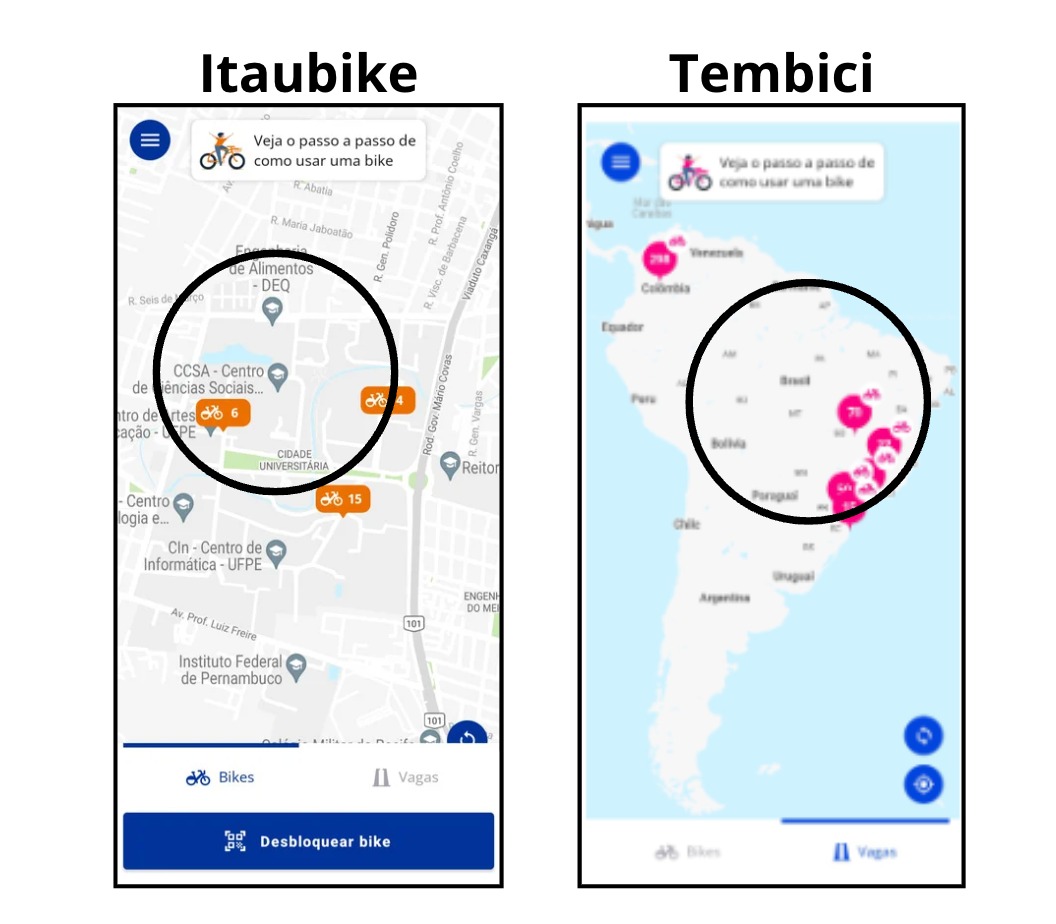}
    \caption{Negative example: Violates the submodules contained in the Engagement Module.}
    \label{front itaubike and tembici}
\end{figure}

Figure \ref{front do tembici and bikeitau} presents an example of a well-recognized interface, where the clickable buttons are clearly visible. In contrast, Figure \ref{front itaubike and tembici} shows a negative example, where it is possible to see "graduation cap" symbols on a map for renting bicycles.

\subsection{Buses}

\begin{figure*}
    \centering
    \includegraphics[width=1\linewidth]{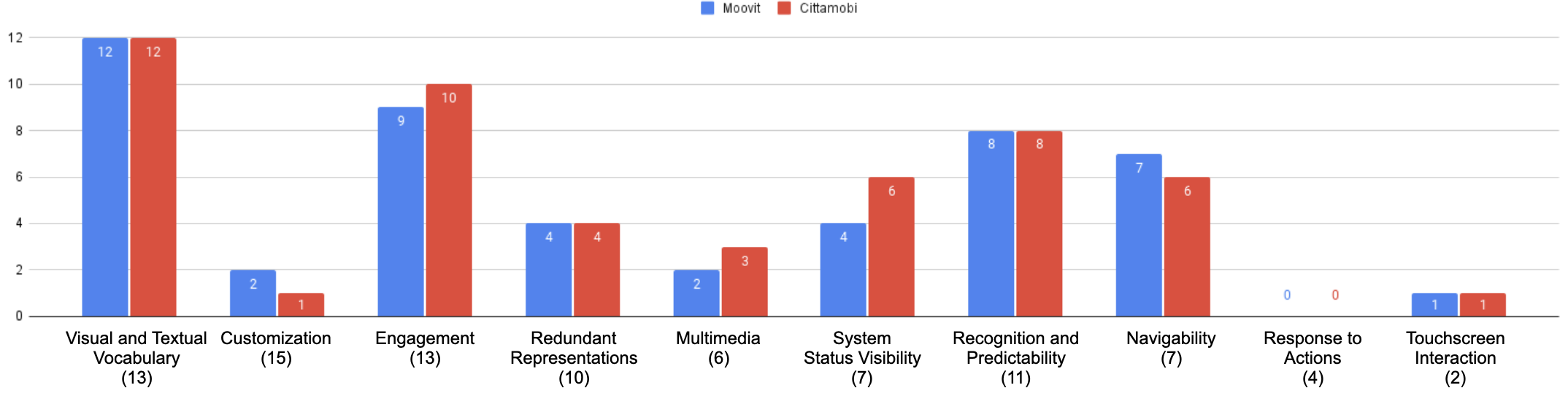}
    \caption{Analysis of the Bus category}
    \label{fig:total-onibus}
\end{figure*}

In the ``Buses'' category, the apps Moovit (version 5.132.0) and Cittamobi (version 7.1.97) were investigated. Similar to the "Bicycles" category, the evaluated apps have some positive points but require improvements to ensure accessibility and an adequate experience for individuals with ASD. Once again, the results show deficiencies in the Customization and Redundant Representations modules, as well as in the Response to Actions module. Figure \ref{fig:total-onibus} presents the comparison of scores between the apps by GAIA modules.

\begin{figure}[h]  
    \centering
    \includegraphics[width=0.5\textwidth]{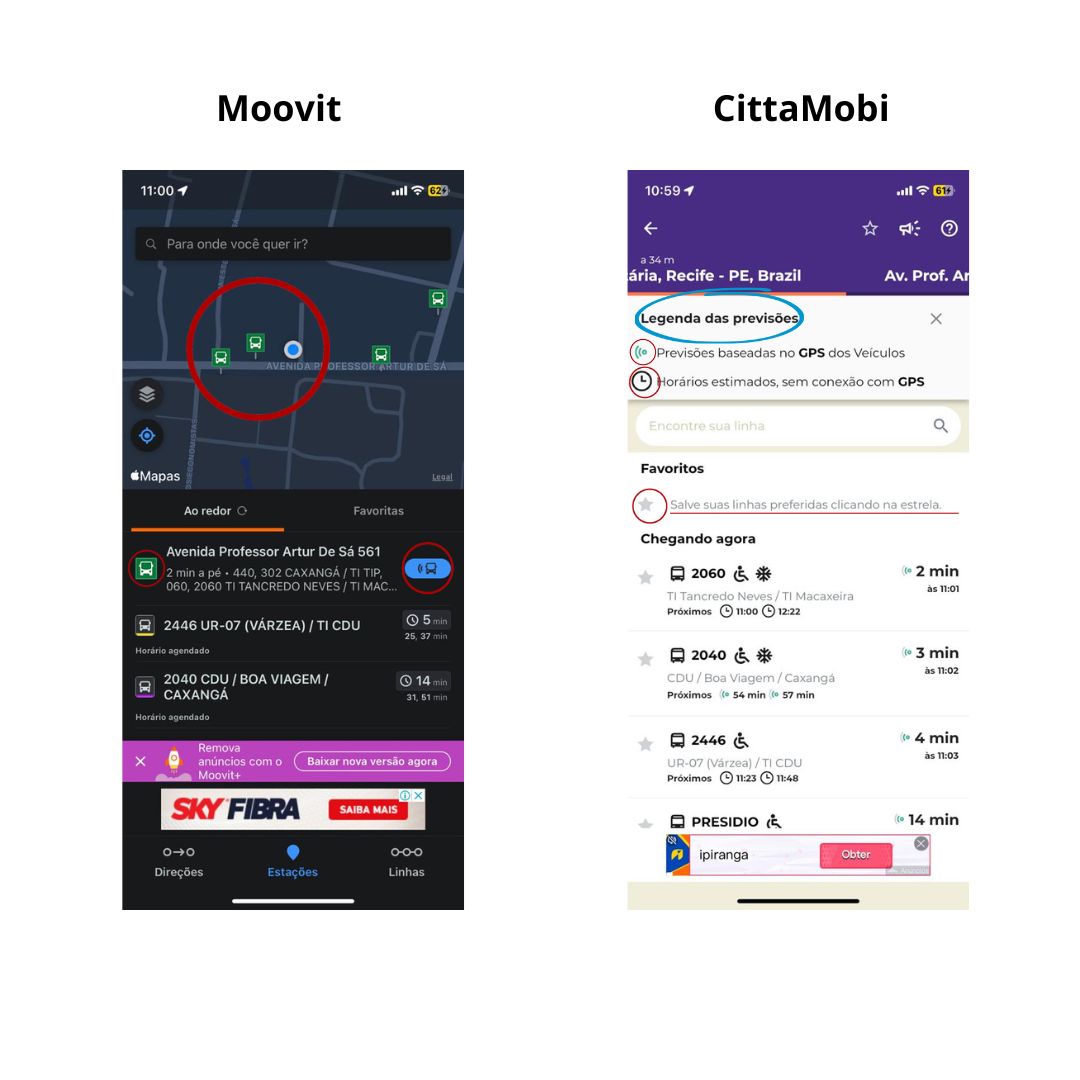}
    \caption{Positive and negative points: Captions in CittaMobi and the lack of them in Moovit.}
    \label{fig:Positivo e Negativo Onibus}
\end{figure}

In the Redundant Representation module, both Moovit and Cittamobi score 4/10, highlighting the lack of representations in visual and audio formats, even though they present symbols and associated names in the interface. There is a significant gap in visual resources, such as images, icons, and sounds, which could facilitate communication. This can be seen in Figure \ref{fig:Positivo e Negativo Onibus}.

In the Response to Actions module, both apps lack feedback for actions performed in the interface, which can be especially challenging for individuals with ASD, given the difficulty in maintaining attention, handling changes, and understanding verbal instructions. Although Cittamobi provides clearer instructions for its activities, its feedback culture is incomplete and may not adequately meet the needs of people with ASD. For instance, the app does not provide clear messages indicating that a certain action should not be performed or how to interact with a specific element; it only offers a functionality to seek help regarding the app’s features.

\begin{figure}[h]  
    \centering
    \includegraphics[width=0.5\textwidth]{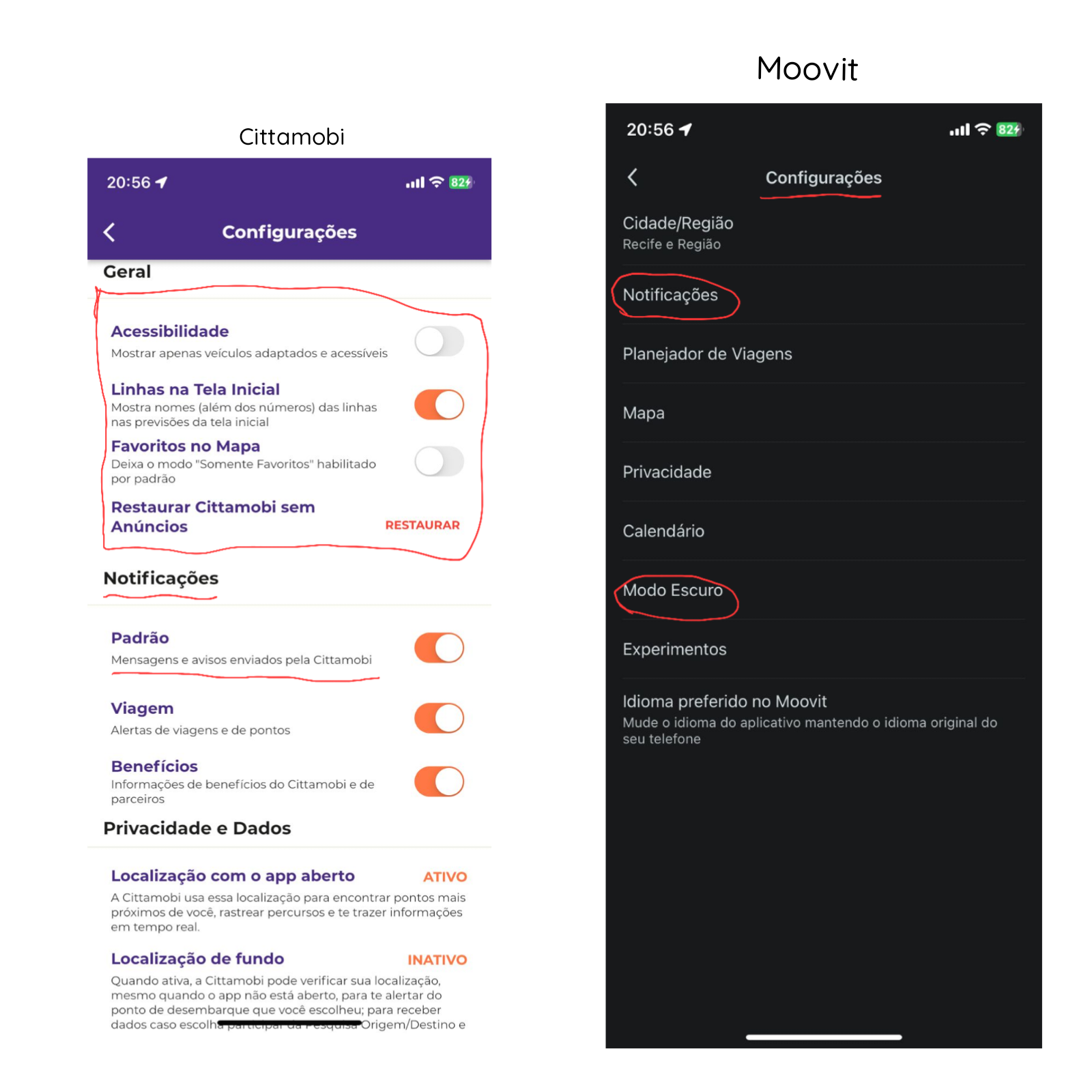}
    \caption{Negative analysis: Visual customization, informational customization, avoid disruptive sounds.}
    \label{fig:personalizacao_onibus}
    
\end{figure}

The lack of customization options, such as font adjustment, varied colors, and audio guidance, is crucial for users with Autism Spectrum Disorder. Few customization options can be observed in Figure \ref{fig:personalizacao_onibus}. 

Regarding the positive points, Moovit stands out in navigability by offering more autonomy through smoother avoidance of automatic redirection to other pages. In the Recognition and Predictability module, Cittamobi provides clearer instructions closer to the elements, while Moovit excels with more predictable behavior when opening new windows, with fewer distractions.

\subsection{Ridesharing}

\begin{figure*}
    \centering
    \includegraphics[width=1\linewidth]{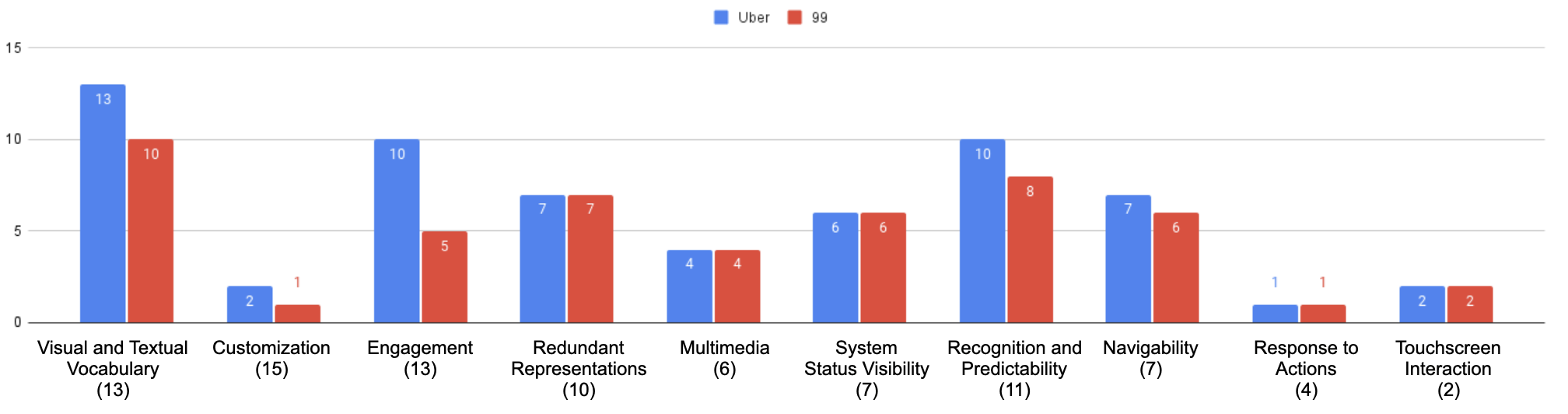}
    \caption{Analysis of the Ridesharing category}
    \label{fig:total-carros}
\end{figure*}

In the ``Ridesharing'' category, Uber (version v3.590.10001) and 99 (version v6.37.4) were analyzed, both of which are among the leading urban mobility companies offering on-demand transportation services across the country. These apps play an important role in urban mobility. However, although this category includes the apps with the highest compliance with GAIA heuristics, both have accessibility limitations. As shown in Figure \ref{fig:total-carros}, the Customization and Response to Actions categories once again appear unsatisfactory in terms of accessibility.

\begin{figure}[h]  
    \centering
    \includegraphics[width=0.45\textwidth]{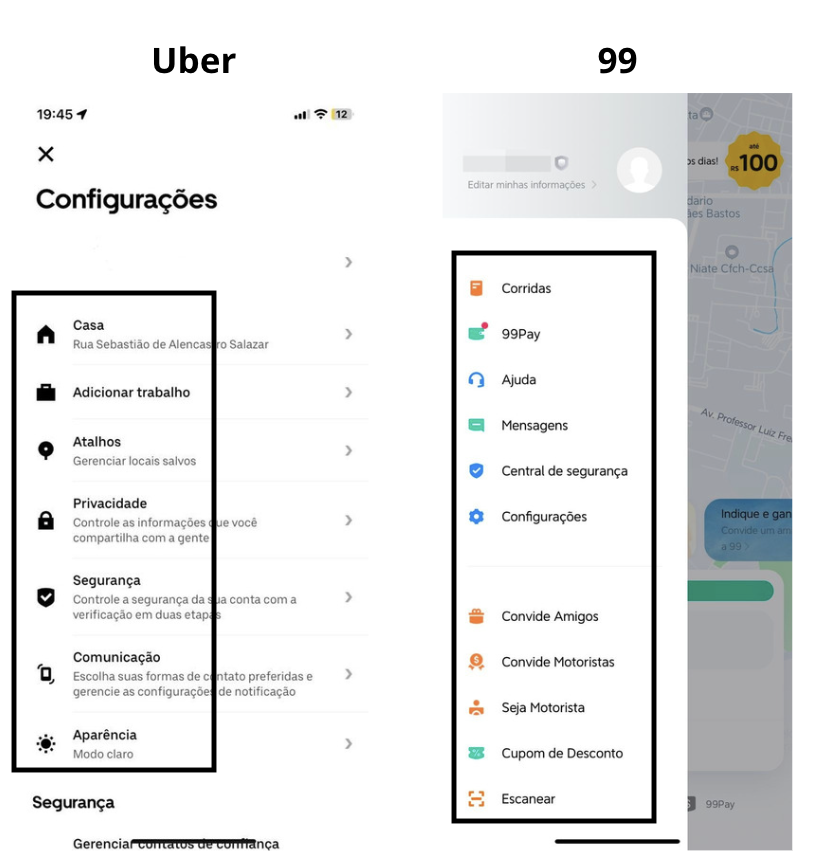}
     \caption{Positive and negative points: Visual Customization, Informational Customization, and Flexible Interfaces submodules in the Customization category.}
     \label{fig:customizar_carros}
\end{figure}

In the Customization category, the car apps mainly fail in visual customization, informational customization, and flexible interfaces. However, they do present icons with captions alongside the text, offering representations in multiple formats. This can be seen in Figure \ref{fig:customizar_carros}.

Regarding the positive aspects, Uber, for example, has an accessible color palette, clearly illustrates its functionalities, and has few elements that may cause distractions. The app's smooth interface allows functionalities to be used intuitively, provides clear information, and makes it easy to find the necessary support to resolve questions and issues. The 99 app offers an accessible color palette, a clear interface, easily understandable information, intuitive functionalities, illustrations that aid usability, accessible support, and organizes information in an intuitive way.

\begin{figure}[h]  
    \centering
    \includegraphics[width=0.5\textwidth]{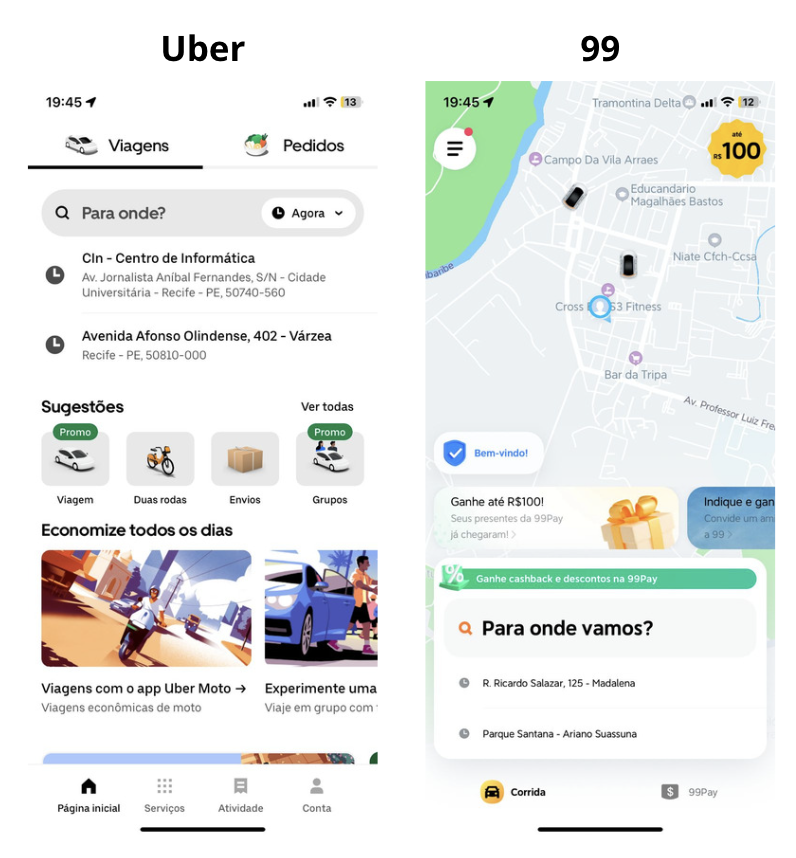}
    \caption{Positive points: Colors, Texts, Provide instructions, Captions, Simple Navigation}
\end{figure}

With the analysis of all categories, it is possible to observe small nuances in each application within its respective category. When comparing the different categories, significant differences can also be noted, even though the apps belong to the same urban mobility sector. In the next section, the discussion of the results will be deepened.

\section{Discussion}

The results showed that, although several modules were unsatisfactory in specific categories, mobility apps in general do not comply with the Customization and Response to Actions modules.

Considering the various GAIA modules, urban mobility apps were found to be insufficient in areas such as customization, redundant representations, multimedia, visibility of system state, response to actions, and interaction with the touchscreen. These deficiencies directly point to the absence of native features associated with accessibility for individuals with Autism Spectrum Disorder. Relating these results to the theory of cognitive overload, it is noted that each individual has a limited capacity to retain and manipulate information in the short term, as proposed by Sweller~\cite{sweller2011cognitive}.

From the perspective of cognitive load theory, intrinsic cognitive load, which is related to the inherent complexity of the task, is exacerbated by the lack of clarity and predictability in the apps, as seen in the cases of Google Maps and Waze, where navigation through multiple icons and information overloads the users' processing capacity. Extraneous cognitive load, which is caused by how information is presented, is also problematic; for example, Moovit and Cittamobi fail to provide immediate and clear feedback, creating noise that hinders understanding of the required actions.

A possible justification for the absence of these tools in the apps would be the argument that smartphones already offer similar features. However, this justification does not hold up when we consider that iOS, for example, offers customization and interaction features with the screen, but there are no guarantees of universal accessibility in all versions of the operating system. Expanding this discussion to Android, it is also clear that its functionalities are not specifically designed to meet the needs of autistic individuals. While some potentially useful features are only available for Pixel smartphones, others are exclusive to newer versions, and the available features of the operating system vary between brands.

The analysis focuses on the needs of individuals with Autism Spectrum Disorder, expanding on the discussion by Araújo et al.~\cite{araujo2011transporte} regarding accessibility, mobility, and quality of life. While that study focused on infrastructure issues impacting general mobility through public transportation, our study provides an understanding of urban mobility applications, addressing accessibility concerns for individuals with ASD in this context. The analysis was expanded to include not only public transportation but also alternative urban travel options, exploring the opportunities available to this specific group and how they are affected during the use of these apps.

The results build on the work of Dattolo and Luccio~\cite{dattolo2017review} by offering a review of guidelines for developing accessible websites and mobile apps for users with ASD, as well as highlighting a lack of quality control and addressing areas for improvement, such as adaptability, language synthesis, and user engagement. It addresses areas for improvement, such as adaptability, language synthesis, and user engagement with ASD individuals.

In this perspective, this study is different in that it focuses on apps for urban mobility, where the operating system was not specifically designed for the accessibility needs of autistic users. Despite this distinction, both apps and websites share similar negative points when it comes to the customization of such software.

It was identified that more specific guidelines for users with ASD are needed, considering their particular challenges, such as limited attention, sensory hypersensitivity, and limited text comprehension. The observations made in this study highlight the lack of customization and more specific guidance, such as alerts in response to incorrect actions. However, this article differs from the work of Dattolo and Luccio~\cite{dattolo2017review}, as the apps analyzed here are not focused on people with ASD, whereas their study focused solely on adult users from this demographic.

A relevant observation is that GAIA was developed with a focus on websites, having limitations related to the evaluation of apps in the mobility context. It is understood that GAIA is a general tool, without a specific context, but it is believed that adaptations may be necessary. For example, one of the only topics in GAIA that addresses navigability is G25 (Simple Navigation), but it is only internal to the app, referring to the breadcrumb trail feature~\cite{lida2003breadcrumb}, which follows this format: Home Page > Section > Subsection; or providing several ways to navigate the site. However, there is nothing that specifically guides the developer on how to create inclusive voice tools that guide an autistic individual when they are moving through the streets and other urban environments. Therefore, it is believed that an adaptation of GAIA is important, given that the urban environment lacks accessibility infrastructure for ASD individuals, such as excessive disruptive sounds, intense light, and high temperatures. This is because, as mentioned by ~\cite{DeslocamentoNoEspaçoUrbano}, there is a fundamental need for Spatial Language for individuals with Autism, due to the aforementioned challenges.

\section{Limitations and Threats}
As with any empirical study, this work has several limitations and threats to validity. In this research, we highlight some important issues.

The first is that a researcher's analysis, even with a guide and training, can be subjective. To address this, the same researcher was assigned to evaluate the same category of apps, and a joint action was carried out with a pilot app among all the evaluators. A possible additional measure to mitigate this limitation would be to use a pair of people for each category of apps.

In addition to the research experience of the evaluators potentially influencing the results, another threat to the validity of this study is the absence of Cohen’s Kappa coefficient calculation to measure agreement between evaluators. Feldt and Magazinius~\cite{feldt2010validity} state that credibility is linked to the confidence that the conclusions are true. To increase confidence, all evaluators had to justify their responses, and a senior researcher reviewed the justifications. Feldt and Magazinius~\cite{feldt2010validity} also state that reliability refers to how consistent the results are, and confirmability refers to how much the findings are shaped by the participants. Furthermore, training and a pilot evaluation were conducted to align agreement and understanding of the guide's use, putting in place a process of response calibration to ensure that all members had a similar interpretation of the guide’s items.

Another limitation refers to the evaluations being restricted to the versions of apps active at the time of evaluation, which may have undergone updates later. Moreover, the results presented here are limited to the categories of the studied apps. The implementation of the operating systems was also not detailed, as each system has some accessibility features. Finally, another important limitation is the absence of autistic individuals in the research team to evaluate the apps. Therefore, the results are concentrated solely on the interpretation of GAIA by the researcher-evaluators.

\section{Conclusion}
This study investigated the degree of compatibility of apps with GAIA, aiming to map possible accessibility issues in urban mobility apps. It was observed that the apps lack a more intuitive approach regarding feedback collection and the availability of clearly identifiable assistance buttons for users with autism. This occurs due to the lack of native accessibility in the apps, as they rely on the adaptability features of mobile devices.

A key contribution of this work is highlighting the importance of reducing the cognitive load in urban mobility apps, which are part of the daily life of autistic individuals, especially those at level 1 of support who have a high degree of independence and autonomy. This study also raises the need to improve accessibility for autistic people. The apps used are crucial in providing greater autonomy for this audience, and if we want to further promote the inclusion of these individuals in society, it is crucial to have apps that are accessible to them. It was noted that, from the perspective of GAIA, the apps present significant limitations in certain categories, such as customization, engagement, response to actions, and redundant representation. On the other hand, other aspects, such as vocabulary and navigability, better meet the heuristics. These results show that technology companies in the urban mobility sector need to be more careful with their apps to make them more accessible to autistic individuals.

We acknowledge that the lack of direct participation of autistic individuals in the app evaluation is a significant limitation. In future research, we aim to involve autistic users in the evaluation and design process to gain more accurate and relevant insights. Additionally, for future work, we suggest adapting GAIA for urban mobility apps so that this type of app can be adequately evaluated in terms of accessibility from the perspective of the autistic population. We also suggest studies to investigate whether cognitive overload or accessibility issues for autistic individuals occur in other widely used apps, such as banking and public service apps.

\begin{acks}
This work was partially supported by INES (\url{http://ines.org.br}), grants CNPq (465614/2014-0) and FACEPE (APQ-0399-1.03/17 and PRONEX APQ/0388-1.03/14).
\end{acks}

\bibliographystyle{ACM-Reference-Format}
\bibliography{references}
\end{document}